\documentclass[a4paper]{iopart}
\usepackage{iopams}     
\usepackage{graphicx}
\usepackage{epsfig}
\usepackage{color}
\usepackage{url}

\usepackage{bm}         
\usepackage{times}
\def\program#1{\textsc{#1}}

\def\etal{{\it et~al.\hbox{}\/}}

\def\Cplusplus{\hbox{C\raise.2ex\hbox{\footnotesize ++}}}

\def\eqref#1{$(\ref{#1})$}

\def\del{\nabla}
\def\P#1{\phantom{#1}}

\def\text#1{\textrm{#1}}

\begin{document}
\title{An explicit harmonic code for black-hole evolution using excision}

\author[Szil\'{a}gyi et al.]{B\'{e}la~Szil\'{a}gyi,$^1$
        Denis~Pollney,$^1$
        Luciano~Rezzolla,$^{1,2}$
        Jonathan~Thornburg$^1$
        and
        Jeffrey~Winicour$^{3,1}$}

\address{$^1$ Max-Planck-Institut f\"ur Gravitationsphysik,
  Albert-Einstein-Institut, Golm, Germany}

\address{$^2$ Department of Physics, Louisiana State University, Baton Rouge, USA}

\address{$^3$ Department of Physics and Astronomy,
             University of Pittsburgh,
             Pittsburgh, USA}


\begin{abstract}
	We describe an explicit in time, finite-difference code designed
	to simulate black holes by using the excision method. The code is
	based upon the harmonic formulation of the Einstein equations and
	incorporates several features regarding the well-posedness and
	numerical stability of the initial-boundary problem for the
	quasilinear wave equation. After a discussion of the equations
	solved and of the techniques employed, we present a series of
	testbeds carried out to validate the code. Such tests range from
	the evolution of isolated black holes to the head-on collision of
	two black holes and then to a binary black hole inspiral and
	merger. Besides assessing the accuracy of the code, the inspiral
	and merger test has revealed that the marginally
	trapped surfaces contained within the common apparent
	horizon of the merged black hole can touch and even intersect.
	This novel feature in the dynamics of the marginally trapped
	surfaces is unexpected but consistent
	with theorems on the properties of these surfaces.
\end{abstract}


\pacs{
     04.25.Dm,  
     02.70.-c,  
     02.70.Bf,  
     02.60.Lj   
     }

%
%

\section{Introduction}
\label{sect-introduction}

The numerical calculation of the inspiral and merger of a binary black
hole system attracted early attention because its dynamic and strongly
relativistic gravitational fields were expected to play a major role in
astrophysics and to provide an excellent arena for studying the fully
nonlinear behavior of gravitation. It has long been
known~\cite{Detweiler-in-Smarr79} that the final decay, coalescence, and
ringdown of such a system is a very strong source of gravitational
radiation.  Because of their complexity and nonlinearity, and the lack of
any continuous symmetries, the Einstein equations cannot be solved
analytically for these systems.  Rather, direct numerical integrations
must be used.

Building on an earlier attempt by Hahn and Lindquist~\cite{Hahn64}, the
first successful numerical simulations of binary black hole systems were
performed in the 1970s by \v{C}ade\v{z}~\cite{Cadez71},
Eppley~\cite{Eppley75}, DeWitt, and their
colleagues~\cite{Smarr76,Smarr77,Smarr79}. These simulations were
restricted to axisymmetry, and used the Arnowitt-Deser-Misner (ADM)
formulation of the Einstein equations~\cite{Arnowitt62}. However, the ADM
equations are now known to be only weakly
hyperbolic~\cite{Frittelli:2000uj}, so they are not suitable for
long-term numerical evolutions despite large-scale
efforts~\cite{Cook97a}.

Starting with the first long-term evolutions by
Pretorius~\cite{Pretorius:2005gq}, there has been remarkable progress in
the simulation of binary black holes. Several groups have since tracked
the inspiral and merger phases using codes solving a conformal-traceless
formulation of the Einstein equations and treating the black holes either
by the excision method~\cite{Diener-etal-2006a} or by the
moving-punctures
method~\cite{Campanelli:2005dd,Baker05a,Gonzales06tr,Herrmann:2006ks}. Since
Pretorius' original work, there has also been substantial progress in
developing mathematical theorems which establish the well-posedness of
the harmonic initial-boundary value problem and the stability of its
finite difference approximations. We have incorporated some of this
theory in developing an explicit in time, finite-difference harmonic code
to treat the excision problem. We present here some promising initial
results indicating that the binary black hole merger can be treated using
excision with a minimal amount of dissipation and with sufficient
accuracy to reveal new and interesting dynamics of the individual
marginally outer trapped surfaces (MOTS) contained within the common
apparent horizon.

Harmonic coordinates were first introduced by
deDonder~\cite{deDonder1921} to reduce Einstein's equations to 10
quasilinear wave equations and they were later extensively developed by
Fock~\cite{Fock-1959} and used by Choquet-Bruhat~\cite{Choquet83} to give
the first well-posed version of the Cauchy problem for the gravitational
field. In a harmonic gauge the spacetime coordinates $x^\mu$ (viewed as a
set of 4~scalar functions) satisfy the curved-space wave equation $\del_c
\del^c x^\mu = 0$. The principle part of the Einstein equations then
reduces to a second-order hyperbolic form or to a first-order symmetric
hyperbolic form, for which there is an extensive mathematical and
computational literature. Many researchers have since implemented
numerical evolution schemes for harmonic formulations of Einstein's
equations~\cite{Garfinkle02,Szilagyi02b,Szilagyi02a,Babiuc:2006wk,Babiuc:2006wp,
Babiuc-etal-2005,Pretorius:2004jg,Pretorius:2005gq,Pretorius:2006tp,
Lindblom:2005qh,Scheel-etal-2006:dual-frame,Palenzuela-etal-2006-boson-stars}
and the related $Z^4$ formulation~\cite{Bona:2002fq,Bona:2004ky}.

The AEI harmonic code presented here has its roots in the Abigel
code~\cite{Szilagyi02a,Babiuc:2006wk,Babiuc-etal-2005}, a second-order
accurate, finite-difference code which incorporates theorems establishing
the well-posedness and numerical stability of the harmonic
initial-boundary value problem. In addition, the AEI code incorporates a
black-hole excision algorithm which allows for motion of the excised
region across the grid, a superluminal evolution algorithm and, except
for regions near the boundaries and for some aspects of the mesh
refinement, a fourth-order accurate finite-difference approximation. It
also utilizes an apparent horizon
finder~\cite{Thornburg95,Thornburg2003:AH-finding}, vertex-centered
mesh-refinement techniques~\cite{Schnetter-etal-03b}, black hole initial
data sets and other features of the \texttt{Cactus} computational
toolkit~\cite{Cactusweb} necessary for black hole simulations. A
Runge-Kutta integration is used to carry out an explicit time evolution.
Note that this differs from the approach of Pretorius who uses a
pointwise Newton-Gauss-Seidel relaxation scheme.

Although some aspects of the code are still under development, most
notably the excision boundary and the outer boundary treatments, we
report here a series of testbeds carried out to validate the code. Such
tests range from the evolution of isolated black holes to the head-on
collision of two black holes and then to a binary black hole inspiral and
merger. While these tests have now become standard, we have found a new
feature in our study of the binary inspiral and merger. In particular,
after a common apparent horizon has formed, our simulations show that the
two individual MOTS approach and continue on to intersect each other.
This novel feature in the dynamics of the MOTS is unexpected but not
unreasonable if these surfaces have to maintain their smoothness as they
meet. It is also consistent with recent theorems concerning the
properties of MOTS (L. Andersson, private communication).

The paper is organized as follows. In
Sect.~\ref{sect-harmonic-system-and-code}, we present the system of
evolution equations and describe its reduction to first-order in time
form, as well as our use of gauge conditions and boundary conditions. In
Sect.~\ref{sect-harmonic/code}, we give a brief discussion of the
numerical techniques we have employed. Section~\ref{sect-harmonic-tests}
collects the results of our tests and the calibration of the code's
accuracy. Finally, Sect.~\ref{sect-discussion} summarizes our results and
the prospects for future work.

\section{Harmonic Evolution System}
\label{sect-harmonic-system-and-code}

\subsection{The evolution system}

Although completely redesigned, the AEI harmonic evolution code is based on the
work presented in Babiuc, Szil\'{a}gyi~\etal~\cite{Szilagyi02a,Babiuc:2006wk,
Babiuc-etal-2005} with modifications to allow for a smooth transition from
subluminal to superluminal evolution, together with higher-order
finite-difference operators and the possibility of excising an arbitrary
portion of the grid (moving black-hole excision).

In a generalized harmonic gauge~\cite{Friedrich96}, the coordinates
$x^\mu=(t,x^i)=(t,x,y,z,)$  satisfy
\begin{equation}
  - \del_a \del^a  x^\mu =\Gamma^\mu  =F^{\mu}
								\;,
\end{equation}
where
\begin{equation}
  \Gamma^\mu :=g^{\rho\sigma} \Gamma^\mu_{\rho\sigma}
 =-\frac{1}{\sqrt{-g}}\partial_\nu {\tilde g}^{\mu\nu}
                                                                    \;,
\end{equation}
with \textit{gauge source functions} $F^\mu(x^\rho,g^{\rho\sigma})$ (which may
depend on the spacetime coordinates and the metric) and with the densitized
4-metric
\begin{equation}
    {\tilde g}^{\mu\nu} := \sqrt{-g} g ^{\mu\nu}
\end{equation}
playing the role of the basic evolution variable. In this harmonic
formulation, the constraints reduce to the gauge condition
\begin{equation}
        C^\mu := \Gamma^\mu - F^\mu =0
	\;,
\label{eqn-harmonic-constraint}
\end{equation}
and the evolution system is based upon the reduced Einstein tensor
\begin{equation}
     E^{\mu\nu}:= G^{\mu\nu} -\nabla^{(\mu}\Gamma^{\nu)}
                 +\frac{1}{2}g^{\mu\nu}\nabla_\alpha \Gamma^\alpha
	             	\;.
 \label{eq:e}
\end{equation}
Here $\Gamma^\nu$ is treated formally as a vector in constructing the
``covariant'' derivative $\nabla^{\mu}\Gamma^{\nu}$. When the constraints
(\ref{eqn-harmonic-constraint}	) are satisfied, this gives rise to a
hyperbolic evolution system
\begin{equation}
     E^{\mu\nu}= -\nabla^{(\mu}F^{\nu)}
                 +\frac{1}{2}g^{\mu\nu}\nabla_\rho F^\rho .
 \label{eq:eeq}
\end{equation}
Provided the gauge source functions do not depend upon derivatives of the
metric, they do not enter the principle part of the system and do not
affect its well-posedness or numerical stability. The evolution system
(\ref{eq:eeq}) takes the specific form of quasilinear wave equations
\begin{eqnarray}
\fl
&\hskip -2.0cm
\partial_\rho \left( g^{\rho\sigma} \partial_\sigma {\tilde g}^{\mu\nu} \right)		
      - 2\sqrt{-g} g^{\rho\sigma} g^{\tau \lambda}
        \Gamma^\mu_{\rho\tau} \Gamma^{\nu}_{\sigma \lambda}
	- \sqrt{-g} (\partial_\rho g^{\rho\sigma}) (\partial_\sigma g^{\mu\nu})
	+ \frac{g^{\rho\sigma}}{\sqrt{-g}} (\partial_\rho g^{\mu\nu})
	        (\partial_\sigma g)
								\nonumber\\
&\hskip -1.0cm
+ \frac{1}{2} g^{\mu\nu}
	  \bigg (
   \frac{g^{\rho\sigma} }{2g\sqrt{-g}} (\partial_\rho g) (\partial_\sigma g)
   + \sqrt{-g} \Gamma^\tau_{\rho\sigma} \partial_\tau g^{\rho\sigma} 
+ \frac{1}{\sqrt{-g}}(\partial_\sigma g)
            \partial_\rho  g^{\rho\sigma}     \bigg )
								\nonumber \\
&\hskip -1.0cm
	+ 2 \sqrt{-g} \del^{(\mu} F^{\nu)}
	- \sqrt{-g} g^{\mu\nu} \del_\rho F^\rho
	+ \sqrt{-g} A^{\mu\nu}
		= 0 \;,
\label{eqn-harmonic-evolution/2nd-order-in-time}
\end{eqnarray}
where we have included the possibility of a constraint-adjustment term
\begin{equation}
A^{\mu\nu}:= C^\rho A^{\mu\nu}_\rho
     (x^\rho,g_{\rho\sigma},\partial_\tau  g_{\rho\sigma})
								\;,
\end{equation}
\textit{i.e.}, a term which vanishes when the constraints are satisfied
and which does not affect the principle part of the evolution system.

Note that we do not explicitly enforce the harmonic
constraints~\eqref{eqn-harmonic-constraint} during the
evolution. Instead, we invoke the Bianchi identities which imply wave
equations of the homogeneous form
\begin{equation}
   g^{\rho\sigma}\partial_\rho \partial_\sigma C^\mu
      + L^{\mu\rho}_\sigma\partial_\rho C^\sigma
       + M^{\mu}_\sigma C^\sigma =0
								\;,
						      \label{eqn-hat-constraint}
\end{equation}
where the matrices ${\boldsymbol L}$ and ${\boldsymbol M}$ are functions
of the metric and its first and second derivatives.

Given constraint-preserving initial and boundary conditions, the
uniqueness of the solutions to (\ref{eqn-hat-constraint}) guarantees
that the harmonic constraints be conserved during the evolution. On the
other hand, constraint-preserving initial data also requires that the
initial Cauchy data satisfy the standard Hamiltonian and momentum
constraints. Also, since the harmonic constraints imply evolution equations
for the lapse and shift, the only remaining free initial data in addition to the
usual Cauchy data (the 3-metric and extrinsic curvature of the Cauchy
hypersurface) are the initial choices of lapse and shift and of the gauge
source functions. Note that an initial choice of the gauge source
functions is effectively equivalent to a choice in the initial evolution
of the lapse and shift.


\subsection{Constraint Adjustment and Damping}

The constraint-adjustments implemented in the code are those investigated
by Babiuc~\etal~\cite{Babiuc-etal-2005} and have the general form
\begin{equation}
\fl \qquad\qquad
A^{\mu\nu}	 := 
	- \frac{a_1}{\sqrt{-g}} C^\rho \partial_\rho {\tilde g}^{\mu\nu}
				  \label{eqn-constraint-adjustment/a1-term}
+ \frac{a_2 C^\rho \del_\rho t}
	        {\varepsilon + e_{\sigma\tau}C^\sigma C^\tau}
		  C^{\mu} C^{\nu}
				   \label{eqn-constraint-adjustment/a2-term}
- \frac{a_3}{\sqrt{-g^{tt}}} C^{(\mu} \del^{\nu)}t \;,
			  \label{eqn-constraint-adjustment/a3-term}
\end{equation} 
where the $a_i>0$ are adjustable parameters (in the runs reported here we
have set $a_i=1$), $e_{\sigma\tau}$ is the natural metric of signature
$({+}{+}{+}{+})$ associated with the Cauchy slicing and $\varepsilon$ is
a small positive number chosen to ensure numerical regularity. The
effects of these adjustments in suppressing long wavelength instabilities
in standardized tests for periodic boundary conditions have been
discussed by Babiuc~\etal~\cite{Babiuc-etal-2005}.

In particular, the first and second terms in the
adjustments~\eqref{eqn-constraint-adjustment/a3-term} have been shown to
be effective in suppressing constraint-violating nonlinear instabilities
in shifted gauge-wave tests. The third term
in~\eqref{eqn-constraint-adjustment/a3-term}, on the other hand, was
first considered in~\cite{Gundlach2005:constraint-damping} and leads to
constraint damping in the linear regime. Although it has been used
effectively by Pretorius~\cite{Pretorius:2005gq,Pretorius:2006tp} in
black-hole simulations, it was not effective in the nonlinear regime of
the shifted gauge-wave test~\cite{Babiuc-etal-2005}.

We also note that the work reported in~\cite{Babiuc-etal-2005} has shown that
adjustments which scale quadratically with  $C^\mu$ (or with higher powers)
take effect too late to counter the growth of a constraint-violating
instability; and this has lead to the specific form for the denominator of the
second term in (\ref{eqn-constraint-adjustment/a3-term}).


\subsection{Gauge Conditions}

As noted earlier, the gauge source functions $F^\mu$ may be chosen to be
arbitrary functions of the spacetime coordinates and metric. They can be
viewed as differential gauge conditions on the densitized metric. This
serves two important purposes. Firstly, it allows for convergence tests
based upon a known spacetime, whose analytic metric $g^{\mu\nu}_{(0)}(x^\rho)$
is specified in a non-harmonic gauge, by choosing
\begin{equation}
F^\mu = - \frac{1}{\sqrt{-g_{(0)}}} \, \partial_\nu {\tilde g}^{\mu\nu}_{(0)}
								\;.
\end{equation}
Using these analytic gauge source functions, in combination with initial
and boundary data consistent with the analytic solution, gives rise to
numerically evolved spacetimes that are identical to the analytic
solution up to discretization error. This is how the convergence tests
reported here have been carried out for the Schwarzschild spacetime
expressed in (non-harmonic) Kerr-Schild coordinates. Secondly, and most
importantly, the gauge source functions can be used to avoid gauge
pathologies.

A major restriction for the form of the gauge source functions is that
they cannot depend on the derivatives of the metric. In particular, they
cannot depend on the location or shape of the MOTS and this
is a problem when moving black holes are present, and where it is
important for the coordinates to be able to ``respond'' to the black hole
motion. In our simulation of binary black holes, we have used the gauge
source function
\begin{equation}
F^\mu =  \frac{\omega}{\sqrt{-g}} \,
	({\tilde g}^{t\mu}-\eta^{t\mu})
                            \;,
\label{eq:gauge-source/qc0}
\end{equation}
where $\eta^{\mu\nu}$ is the Minkowski metric and where $\omega =
\omega(x^i)$ is a smooth, spherically symmetric, time-independent
weighting function with $\omega=1$ over most of the computational domain,
but with $\omega=0$ in some neighborhood of the outer boundary. When
spatial derivatives are neglected and $\omega=1$, the resulting gauge
condition takes the simpler form
\begin{equation}
 \partial_t {\tilde g}^{t\mu}=-({\tilde g}^{t\mu}-\eta^{t\nu})\;,
\end{equation}
showing that it forces the densitized lapse and shift to relax to their
Minkowski values. 

In our first attempts at binary black hole simulations, we have found
that this choice of gauge source function keeps the lapse and shift under
reasonable control. Similar choices of gauge source functions have been
used with success in other binary black hole
simulations~\cite{Pretorius:2005gq,Pretorius:2006tp}.


\subsection{Boundary Conditions}


Our evolution domain has a timelike outer boundary and a smooth,
spacelike excision boundary inside each MOTS. The harmonic
evolution system, in the second-order form
(\ref{eqn-harmonic-evolution/2nd-order-in-time}), consists of quasilinear
wave equations whose characteristics are identical to the null directions
determined by the metric. As a result, all characteristics leave the
spacelike excision boundaries and no boundary conditions are necessary
(or allowed).

At the timelike outer boundary, any dissipative boundary condition for
the wave equation with shift leads to a well-posed initial-boundary value
problem (IBVP). Such dissipative boundary conditions were first worked
out in the one-dimensional (1D)
case~\cite{Calabrese:2005sec,Szilagyi05,Calabrese:2005fp} and general
results for the 3D case have been discussed recently in
~\cite{Babiuc:2006wk,Motamed06}. For a boundary with normal in the $+x$
direction, such dissipative boundary conditions have the form
\begin{equation}
\left [ (1-\kappa ) \partial_t + \kappa g^{z\rho}\partial_\rho \right] 
{\tilde g}^{\mu\nu}
	= q^{\mu\nu}, \qquad 0\le \kappa \le 1 ,
\label{eq:dissbc}
\end{equation}
for each component ${\tilde g}^{\mu\nu}$, where $q^{\mu\nu}$ is the
boundary data. The choice $\kappa = 0$ gives a Dirichlet condition and
$\kappa =1$ gives a Neumann condition. Dirichlet and Neumann conditions
are marginally dissipative in the sense that they are purely reflective
for modes with $ q^{\mu\nu}=0$. A strictly dissipative Sommerfeld-type
condition arises when $\kappa$ is chosen so that the derivative in the
left hand side of (\ref{eq:dissbc}) lies in the outgoing null direction.

In order for the IBVP to be constraint preserving, the boundary data
$q^{\mu\nu}$ must be assigned to enforce a homogeneous, dissipative
boundary condition on the constraints $C^\mu$. Then, with proper
initialization, the uniqueness of solutions to
eqs.~(\ref{eqn-hat-constraint}) ensures that the constraints are
satisfied throughout the evolution.  The first proposal for such
constraint-preserving boundary conditions for the harmonic system
consisted of a combination of 3 Dirichlet and 7 Neumann conditions on the
components of ${\tilde g}^{\mu\nu}$~\cite{Szilagyi02a}. However,
numerical studies~\cite{Babiuc:2006wk} showed that these
Dirichlet-Neumann boundary conditions were effective in carrying the
signal off the grid but that their marginally dissipative nature
reflected the noise and gave poor results in highly nonlinear tests.

The first example of strictly dissipative constraint-preserving boundary
conditions which would in principle allow numerical error to leave the
grid, was given for a tetrad formulation of the Einstein equations by
Friedrich and Nagy~\cite{Friedrich99}. Constraint-preserving boundary
conditions of the Sommerfeld type which lead to a well-posed IBVP for the
non-linear harmonic formulation have subsequently been
formulated~\cite{Kreiss-Winicour-2002}. These Sommerfeld-type boundary
conditions have been incorporated in a numerical code which gives vastly
superior performance in nonlinear test problems than the
Dirichlet-Neumann scheme~\cite{Babiuc:2006wp}. However, we have not yet
implemented these condition here; instead we have used a naive version of
Sommerfeld boundary conditions which not only is not
constraint-preserving but whose numerical implementation is only
second-order accurate. In addition, as explained further in
Sect.~\ref{sect-harmonic/code}, the code uses summation-by-parts (SBP)
difference operators which are fourth-order accurate in the interior but
only second-order accurate in the vicinity of the outer boundary.


\subsection{Reduction to First-Order in Time}

In contrast to what is done in~\cite{Pretorius:2005gq,Pretorius:2006tp},
where the harmonic evolution system is second-order in time, we find it
convenient to discretize and use the method of lines to time-integrate an
evolution system which is first-order in time. We note that the reduction
to first-order in time can be done in a number of ways, some of which may
have very different stability properties when discretized
(see~\ref{app-summation-by-parts} for a discussion).

Here, we introduce the auxiliary variables
\begin{equation}
\hat{Q}^{\mu\nu} := g^{tt} \partial_t \tilde{g}^{\mu\nu}
                + w g^{ti} \partial_i \tilde{g}^{\mu\nu}
								\;,
\label{eqn-Qhat-defn}
\end{equation}
where $w = w(x^i)$ is a smooth weighting function with $w=1$ over most of the
computational domain but with $w=0$ in a neighborhood of the outer boundary. We
rewrite the definition~\eqref{eqn-Qhat-defn} to obtain the time derivatives of
$\tilde{g}^{\mu\nu}$ in terms of $\hat{Q}^{\mu\nu}$ and spatial derivatives of
$\tilde{g}^{\mu\nu}$
\begin{equation}
\partial_t \tilde{g}^{\mu\nu}
	= \frac{1}{g^{tt}} \left(
	   \hat{Q}^{\mu\nu} - w g^{ti} \partial_i   \tilde{g}^{\mu\nu}
			   \right)
								\;.
	  \label{eqn-dt-gab=fn(Qhat-and-di-gab)}
\end{equation}
Next, we use the identity~\eqref{eqn-dt-gab=fn(Qhat-and-di-gab)} to
re-express the principle part of the harmonic evolution
equations~\eqref{eqn-harmonic-evolution/2nd-order-in-time} as
\begin{equation}
\fl \quad
\partial_\rho \left( g^{\rho\sigma} \partial_\sigma {\tilde g}^{\mu\nu} \right)
  =\partial_t \hat{Q}^{\mu\nu}
      + (1-w) \partial_t(g^{ti}\partial_i \tilde{g}^{\mu\nu}) 
      + \partial_i(g^{it}\partial_t {\tilde g}^{\mu\nu})
      +\partial_i(g^{ij}\partial_j {\tilde g}^{\mu\nu}) \;,
\label{eq:ppreduct}
\end{equation}
and after using~\eqref{eqn-dt-gab=fn(Qhat-and-di-gab)} to convert all
time derivatives of $\tilde{g}^{\mu\nu}$ in~\eqref{eq:ppreduct}
and~\eqref{eqn-harmonic-evolution/2nd-order-in-time} into spatial
derivatives, we obtain an equation of the form
\begin{equation}
\partial_t \hat{Q}^{\mu\nu}
	= F^{\mu\nu}(\tilde{g},
	    \partial_i \tilde{g},
	    \partial_{ij} \tilde{g},
	    \hat{Q},
	    \partial_i \hat{Q})
								\;.
\label{eqn-Q-hat-evolution}
\end{equation}
Equations~\eqref{eqn-dt-gab=fn(Qhat-and-di-gab)}
and~\eqref{eqn-Q-hat-evolution} represent our basic evolution equations
for the field variables $\tilde{g}^{\mu\nu}$ and $\hat{Q}^{\mu\nu}$,
respectively.

\section{Numerical Implementation}
\label{sect-harmonic/code}

\subsection{Finite-difference algorithms}

The code solves the finite-difference equations on a Cartesian grid with
finest resolution $\Delta x^i=h$, using a cubic outer boundary and with
excision boundaries for each black hole. Vertex-centered mesh
refinement is applied using the \texttt{Carpet} driver
~\cite{Schnetter-etal-03b}, within the framework of the \texttt{Cactus}
computational toolkit~\cite{Cactusweb}. The time evolution is carried out
by the method of lines using a fourth-order Runge-Kutta scheme, with a
fifth-order spatial prolongation and a second-order time interpolation to
provide fine-grid boundary data at mesh-refinement boundaries. 

While the bulk of the code uses fourth-order accurate centered difference
operators to approximate spatial derivatives, in a neighborhood of the
outer boundary we approximate the spatial derivatives by diagonal norm
SBP difference operators of fourth-order interior accuracy and of
second-order accuracy at the boundary. More specifically, on a grid $x_I
= x_0 + i h$ with boundary at $x_0$, these operators, as described by
Mattsson and Nordstr{\"o}m
~\cite{Mattsson-Nordstrom-2005:SBP-operators}, are
\begin{eqnarray}
&&\hskip -1.0cm
(\partial_x f)_{i=1} \rightarrow 
 \frac{1}{h} \left( \frac{1}{2} f_{[2]} - \frac{1}{2} f_{[0]} \right)\;,
\\
&&\hskip -1.0cm
(\partial_x f)_{i=2} \rightarrow 
 \frac{1}{h} \left(
- \frac{4}{43} f_{[4]}
+ \frac{59}{86} f_{[3]}
- \frac{59}{86} f_{[1]}
+ \frac{4}{43} f_{[0]}
\right)\;, 
\\
&&\hskip -1.0cm
(\partial_x f)_{i=3} \rightarrow 
 \frac{1}{h} \left(
- \frac{4}{49} f_{[5]}
+ \frac{32}{49} f_{[4]}
- \frac{59}{98} f_{[2]}
+ \frac{3}{98} f_{[0]}
\right)\;,
\end{eqnarray}
and
\begin{eqnarray}
&&\hskip -1.0cm
(\partial_x^2 f)_{i=1} \rightarrow 
 \frac{1}{h^2} \left( f_{[2]} - 2 f_{[1]} + f_{[0]} \right)\;,
\\
&&\hskip -1.0cm
(\partial_x^2 f)_{i=2} \rightarrow 
 \frac{1}{h^2} \left(
- \frac{4}{43} f_{[4]}
+ \frac{59}{43} f_{[3]}
- \frac{110}{43} f_{[2]}
+ \frac{59}{43} f_{[1]}
- \frac{4}{43} f_{[0]}
\right)\;,
\\
&&\hskip -1.0cm
(\partial_x^2 f)_{i=3} \rightarrow 
 \frac{1}{h^2} \left(
- \frac{4}{49} f_{[5]}
+ \frac{64}{49} f_{[4]}
- \frac{118}{49} f_{[3]}
+ \frac{59}{49} f_{[2]}
- \frac{1}{49} f_{[0]}
\right) \;.
\end{eqnarray}

For grid points at the outer boundary, on the other hand, all components
of $\hat Q^{\mu\nu}$ are updated using a flat-spacetime, homogeneous
Sommerfeld boundary condition. With the outer boundary located in the
weak-field regime, a Sommerfeld condition applied to $\hat Q^{\mu\nu}$ 
is equivalent to setting the Sommerfeld derivative of $\tilde g^{\mu\nu}$
to the value of this Sommerfeld derivative at $t=0$, as determined
by the initial data.  (By implication, our boundary algorithm is compatible
with the initial data.) This boundary condition is effective in maintaining
numerical stability. However, it is not constraint-preserving and
is a prime target for future code improvement.

In the part of the computational domain near the outer boundary where $w=0$,
the evolution algorithm for ${{\hat Q}}$ and ${{\tilde g}}$  
 reduces to a fourth-order version of the subluminal evolution algorithm
for evolving the wave equation with shift discussed in~\cite{Motamed06}. This
algorithm is known to be unstable in the region where the shift is superluminal
(e.g. near the excision boundary). In this superluminal region, we set $w=1$ so
that the  definition of ${{\hat Q}}$ 
 [cf. eq.~\ref{eqn-Qhat-defn}]
makes ${{\hat Q}}$ the derivative of ${{\tilde g}}$ in
the normal direction to the Cauchy hypersurfaces, which stabilizes the
algorithm. A more detailed discussion of this subluminal-superluminal blending
is discussed in~\ref{app-summation-by-parts}.

Another important ingredient of our code is numerical dissipation.
We find this essential in keeping  the algorithm stable in the neighborhood
of the excision domain.  In addition, this is also helpful in killing
off high-frequency noise generated at the mesh-refinement boundaries.
In the interior of the grid,
numerical dissipation is added at $O(h^5)$ in the form
\begin{equation}
  \partial_t^2 \tilde g^{\mu\nu} \rightarrow  \partial_t^2 \tilde g^{\mu\nu}
     +\frac{1}{64}h^5 \sum_i \epsilon_i (D_{+i} D_{-i})^3 
	\partial_t \tilde g^{\mu\nu}
                                         \;,
					 \label{eq:dissop}
\end{equation}
where $D_{\pm i}$ are the forward and backward difference operators in
the $x^i$ direction and $\epsilon_i$ is a smooth weighting function. In
the neighborhood of a face of the outer boundary with normal in the
$x$-direction, we set $\epsilon_x =0$ so that the dissipation applies
only in the tangential directions. In carrying out convergence tests for
a Schwarzschild black hole, we choose $\epsilon_i=0.2$ outside the
apparent horizon (AH) and $\epsilon_i=2$ inside the AH (except for a
transition region). In the two black hole simulations, we choose
$\epsilon_i=1$ outside the AH and $\epsilon_i=2$ inside the AH.

\subsection{Moving Excision}

The excision algorithm is driven by the apparent horizon finder algorithm
~\cite{Thornburg95,Thornburg2003:AH-finding}. Strictly speaking, this algorithm
searches for MOTS, regardless of whether these are apparent horizons or not. We
recall that MOTS are defined as smooth, compact 2-dimensional surfaces whose
outgoing normal null geodesics have zero expansion. With respect to a $3+1$
foliation, the apparent horizon is defined as the 3-dimensional hypersurface
traced out by the outer boundary of the trapped region in each time slice. If
sufficiently smooth, the apparent horizon is foliated by MOTS.

A smooth spacelike boundary is used to excise a region
inside each MOTS, resulting in a jagged boundary in the Cartesian grid. The
excision boundary is chosen to be centered inside the MOTS and scaled in
coordinate size to be $0.7$ the size of the MOTS in the binary simulations
and $0.8$ in the Schwarzschild black hole tests. 

We keep the same interior evolution stencil near the excision boundary by
introducing the necessary ghost points. Because the dissipation
operator (\ref{eq:dissop}) would require an excessive number of ghost
points we replace it with a third-order form $h^3 (D_{+i} D_{-i})^2$ near
the excision boundary. Values at the required ghost points are supplied
by an extrapolation scheme which was proved to be stable for the case of
a boundary aligned with the grid~\cite{Motamed06}. We have generalized
this to the case of a generic smooth boundary in a Cartesian grid following
the ``embedded-boundary'' method developed by Kreiss and
Petersson~\cite{Kreiss04a} for formulating a stable Neumann
condition. More specifically, we construct a vector $v^i$ by taking the
flat-space displacement from the centroid of the excised region to the
current position, and require that
\begin{equation}
     \sum_i( v^i D_{\pm i})^3 \tilde g^{\mu\nu}= 0
\quad \text{and} \quad
     \sum_i( v^i D_{\pm i})^3 \hat Q^{\mu\nu}= 0
                                          \;, 
     \label{eq:extrap}
\end{equation}
where the one-sided differences $D_{\pm i}$ correspond to the sign of $v^i$.
The extrapolation condition (\ref{eq:extrap}) is applied iteratively at the
points near the boundary until the full stencil of ghost points is updated.

Extra care needs to be paid to the identification of ghost points when the
excision domain is moving across the grid. In particular, grid-points that
become interior  points at $t^{N} := t_0 + N \Delta t$ but were ghost
points at $t^{N-1}$ need to be treated as ghost points, \textit{i.e.}, the excision
algorithm must fill them with extrapolated values during the time-integration
from $t^{N-1}$ to $t^{N}$.  However, when evolving from $t^{N}$ to $t^{N+1}$,
these same grid-points need no longer be treated as ghost points and can then
be labeled as evolution points.

\section{Code Tests}
\label{sect-harmonic-tests}

We now present a series of tests to validate our code's stability and
accuracy, starting with the evolution of Schwarzschild black holes using
static gauge source functions, followed by the head-on collision of
two equal-mass black holes and finally the inspiral and merger of the
QC-0 initial data set for binary black holes.

\begin{figure}
\begin{center}
\includegraphics[width=0.92\textwidth]{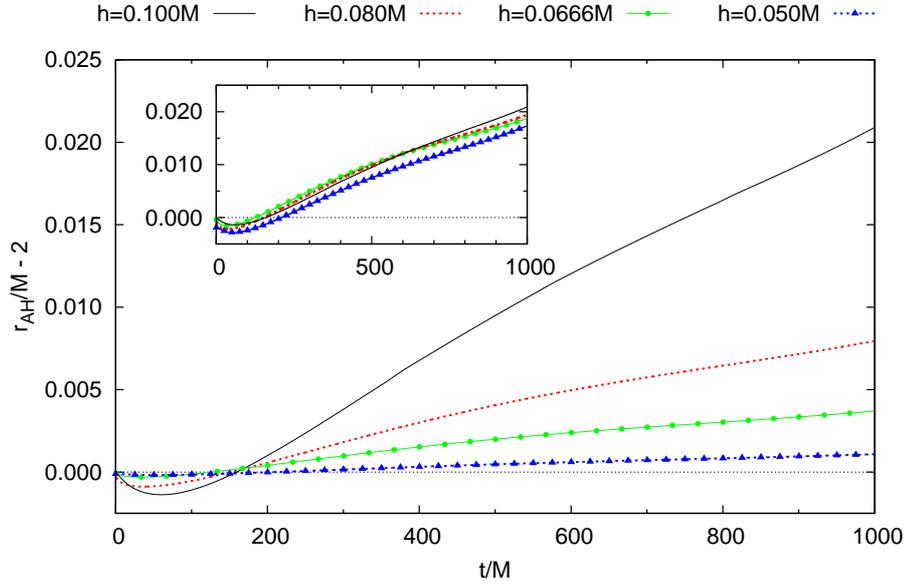}
\end{center}
\caption{Error $r_{_{\rm AH}}/M - 2$ in the areal radius of the apparent
  horizon as a function of time for evolutions of Schwarzschild initial
  data at 4~different resolutions.  The main figure shows $r_{_{\rm AH}}/M - 2$,
  while the inset shows $(0.1M/h)^4 \times (r_{_{\rm AH}}/M - 2)$, demonstrating
  that the error shows fourth-order convergence to zero as the resolution
  is increased.  }
\label{fig-bh-radius-of-time}
\end{figure}

\begin{figure}
\begin{center}
\includegraphics[width=0.92\textwidth]{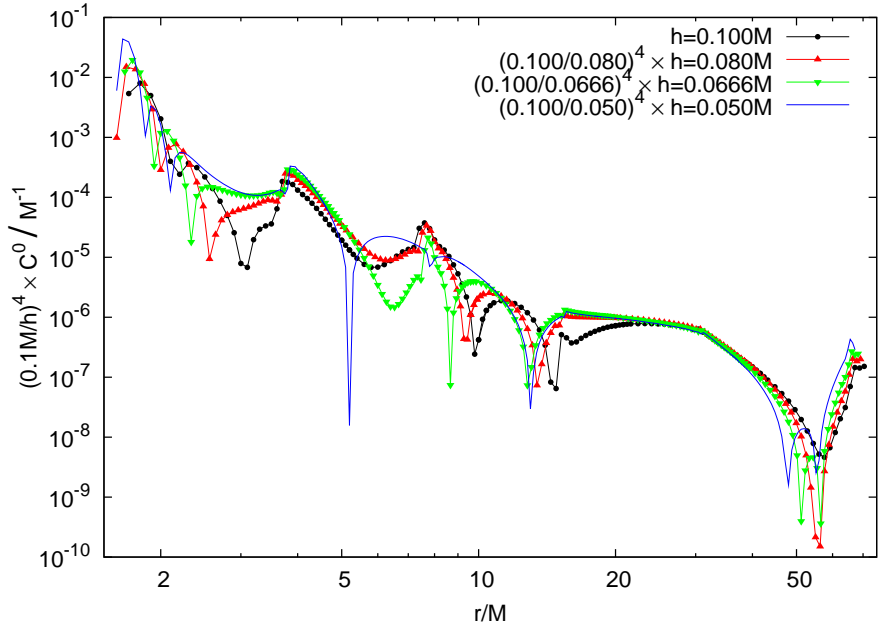}
\end{center}
\caption{Scaled harmonic constraint $(0.1M/h)^4 \times C^0$ at $t=200M$
	along the $x$~axis, for evolutions of Schwarzschild initial data
	at 4~different resolutions.  }
\label{fig-Schw-constraint-on-xaxis}
\end{figure}

\subsection{Single Black Holes}

We have evolved Schwarzschild black holes for long periods of time to
test the stability and accuracy of our evolution algorithm, excision
scheme and outer boundary treatment. To study the code's convergence with
resolution, we made evolutions with the finest grid spacing set to
$h=0.100M$, $0.080M$, $0.0666M$, and $0.050M$. Each evolution used a similar
grid structure, with 5~levels of nested 2:1~mesh refinement with the
finest grid (spacing $h$) extending from the origin to $4M + 9\,h$, the
next coarsest (spacing $2\,h$) from the origin to $8M + 9\,h$ and so on,
up to the coarsest grid (spacing $16\,h$) from the origin to $64M$.  The
tests were carried out in octant symmetry. For representative runs, we
cross-checked that identical output was produced in full-space
simulations.

Each evolution was run for $1000M$, with no sign of instability for this full
duration.  We measure the accuracy of these evolutions by monitoring the
apparent horizon's areal radius and the harmonic constraint~$C^0$. (Convergence
results for the error in $\tilde{g}^{00}$ are very similar to those for $C^0$.)

Figure~\ref{fig-bh-radius-of-time} shows the error in the apparent
horizon areal radius $r_{_{\rm AH}}/M - 2$ for these evolutions.  After
an initial transient, each evolution displays a slow growth (roughly
linear in time) in the apparent horizon. The growth rates are small, and
show fourth-order convergence to zero with the resolution.
Figure~\ref{fig-Schw-constraint-on-xaxis}, on the other hand, shows the
magnitude of $C^0$ along the $x$ axis at $t=200M$. It is clear that while
there are regions of excellent pointwise fourth-order convergence, there
are also regions where mesh-refinement and other effects make the
convergence more problematic.

\begin{figure}
\begin{center}
\includegraphics[width=0.92\textwidth]{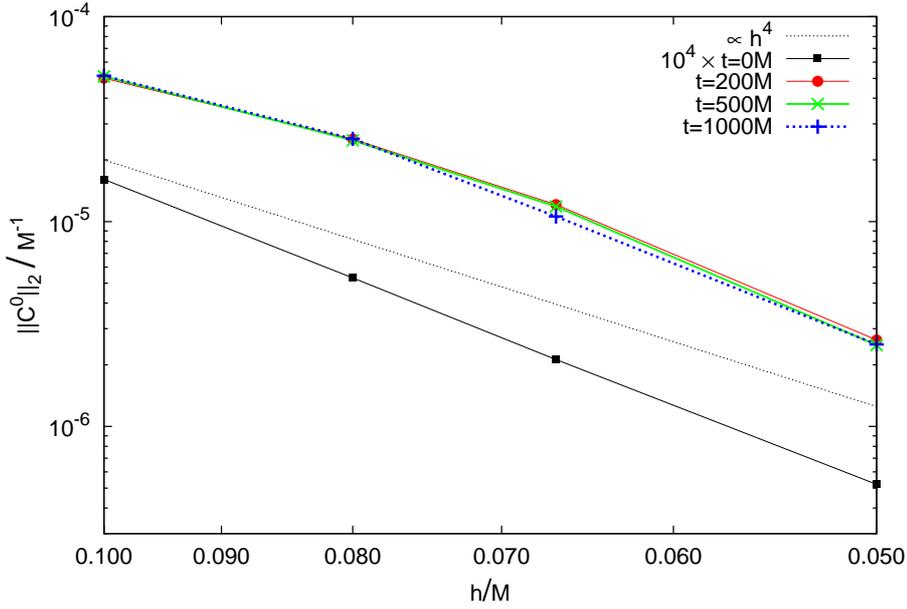}
\end{center}
\caption{Masked 2-norm of the harmonic constraint $\| C^0\|_2$ as a
	function of the finest grid resolution $h$, at selected times,
	for evolutions of Schwarzschild initial data at 4~different
	resolutions.  Note that the $t=200M$, $t=500M$, and $t=1000M$
	curves are almost coincident (because the norms are almost
	time-independent over this range of times).  }
\label{fig-Schw-constraint-norms-of-dx}
\end{figure}

To gain a clearer picture of the overall convergence behavior of these
evolutions and because the error is by far and large dominated by a small set
of grid points near the excision boundary, we use a ``masked'' 2-norm $\|
C^0\|_2$ defined as the root-mean-square norm of $C^0$ over only those grid
points which are outside the apparent horizon and do not have any
finer--refinement-level grid points overlaying them.  For example, all of the
finest grid outside the apparent horizon would be included in the masked norm,
but only that part of the grid with spacing $2\,h$ which is outside
$4M{+}9\,h$ would be included, and similarly for the other coarser refinement
levels.

After a short initial transient (lasting a time comparable to the
outer-boundary crossing time), $\| C^0\|_2$ is essentially
time-independent in each run.
Figure~\ref{fig-Schw-constraint-norms-of-dx} shows the variation of $\|
C^0\|_2$ with resolution at selected times.  As can be seen, the norm
decreases with resolution, with a convergence exponent which is always
$\ge 3$, but varies with resolution.  We attribute this behavior to not
having sufficient resolution to see the asymptotic convergence behavior
as introduced by the dissipation operator.

Similar results have also been obtained for spinning black holes, which we have
evolved with very high spins. Simulations of Kerr black-holes with spins
$J/M^2$ up to $0.99$ have shown no apparent signs of instability up to $t = 100
M$. We have not yet investigated them further.


\subsection{Binary Black Hole Mergers}

\subsubsection{Head-on collision}

The binary black hole collision we have carried out is an equal-mass,
non-spinning head-on collision of two Brill-Lindquist black holes of
masses $0.5$ each, located on the $z$ axis at $z = \pm 1.16$.

The test was carried out in octant symmetry, using a pure harmonic gauge
(\textit{i.e.},~$F^{\mu}=0$). The numerical grid was set up using nine levels of
refinement, with grid-step $h^{(n)} = 3.2 \times 2^{-n}, n = 0,\,
\ldots,\, 8$ giving a grid-step of $h = 0.0125$ on the finest refinement
level and an outer boundary at $144 M$. The chosen grid setup is such that the
initial black holes are contained within the bounding box of the finest
grid and hence no motion of the finer grids is needed.

\begin{figure}[tbp]
\begin{center}
\includegraphics[width=0.92\textwidth]{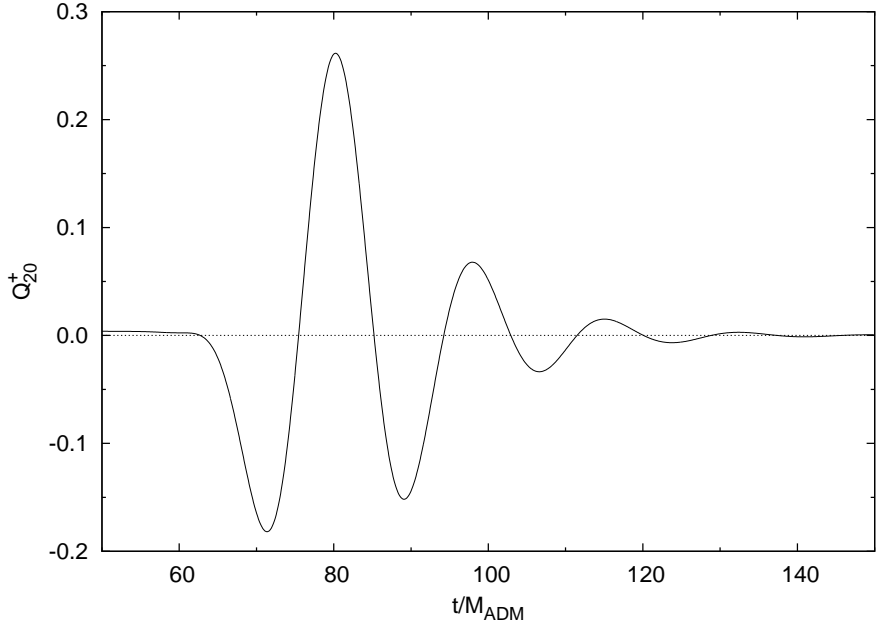}
\end{center}
\caption{ Zerilli waveform for the head-on problem, extracted at $R=60M$.
The outer boundary in this test was at $L=144 M$.}
\label{fig-headon-waveform}
\end{figure}

After the merger, when the final horizon has a an ellipsoidal shape with
maximum and minimum radii in a ratio $r_{min}/r_{max} \gtrsim 0.6$, the
finest refinement level is dropped. This is justified by the fact that
the coordinate-radius of the final apparent horizon radius is more than
double that of the two individual MOTS.  Overall, the simulation shows no
sign of instability and Fig.~\ref{fig-headon-waveform} shows the
$\ell=2$, $m=2$, even-parity multipole of the Zerilli function $Q^+_{20}$
as extracted at $R=60M$, indicating a well captured quasi-normal ringing
of the black hole.

\subsubsection{QC-0 inspiral}

\begin{figure}
\begin{center}
\includegraphics[width=0.8\textwidth]{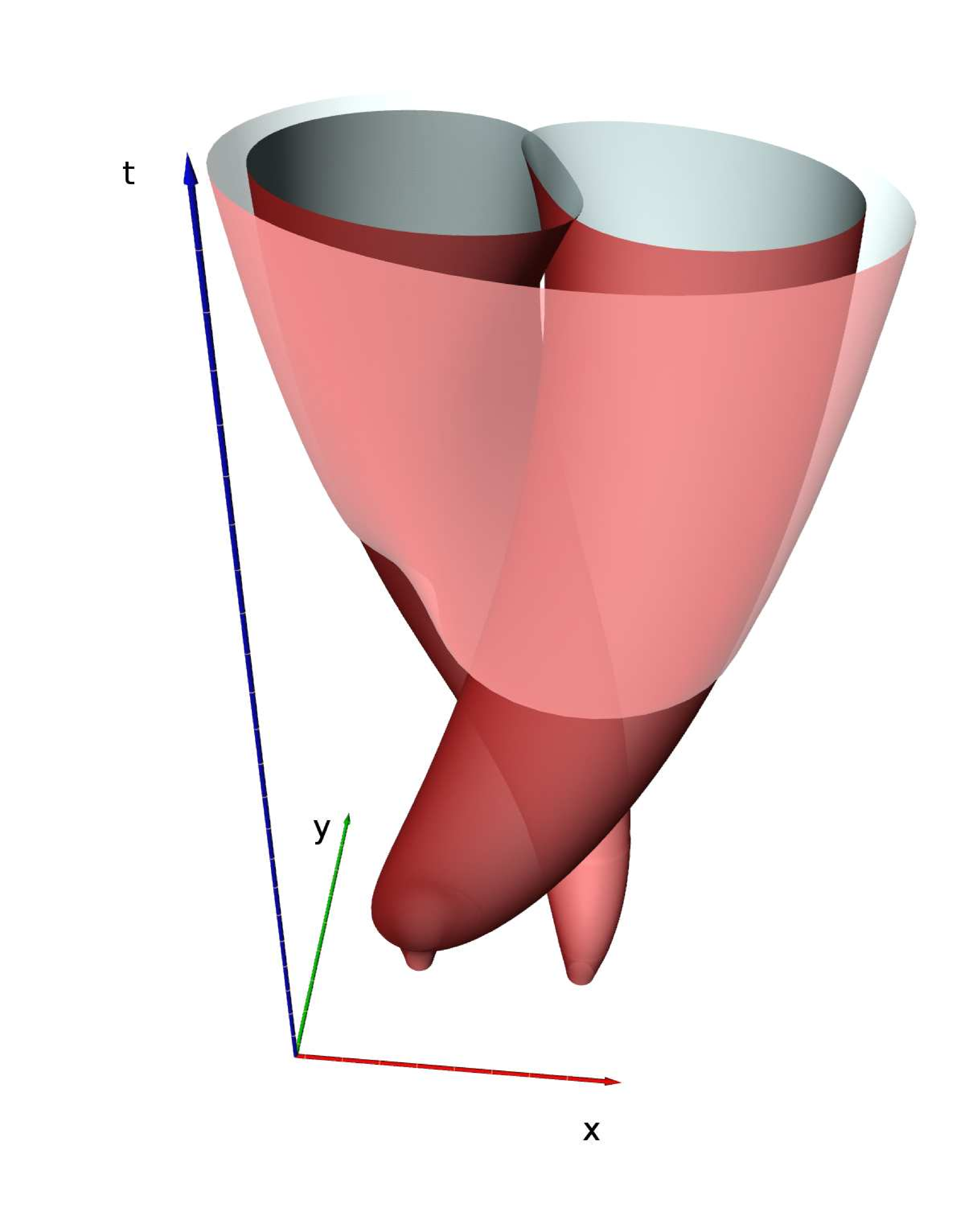}
\end{center}
\caption{
	are $(x,y,t)$. The overlap of the MOTS
	 is clearly visible.
	Note that these surfaces are actually smooth everywhere;
	their apparent non-smoothness in some parts of this figure is an
	artifact of the perspective projection.  }
\label{fig-qc0-PairOfPants-3D}
\end{figure}

We next consider the evolution of the QC-0 initial data defined
by~\cite{Baker:2002qf} and constructed with the highly-accurate
spectral elliptic solver of~\cite{Ansorg:2005bp}. In contrast with what
was done for the head-on collision, we here use the gauge source
function (\ref{eq:gauge-source/qc0}) in order to keep the lapse from
collapsing and the shift from forming large coordinate distortions in the
strong-field region. While helping in terms of stability, these gauge
source functions also lead to a coordinate-growth of the MOTS.

The simulation started out with nine levels of mesh refinements, with the
spacing on the individual refinement levels being $h^{(n)} = 2.048 \times
2^{-n}, n = 0,\, \ldots,\, 8$ and yielding a resolution of $0.008M$ on
the finest refinement level, with an outer boundary at $\approx 160 M$.
The simulation was carried out in the $x \geq 0, z \geq 0$ quadrant,
taking advantage of the reflection symmetry across $z=0$ and the $\pi/2$
rotation symmetry around the $z$ axis.  Around each black hole we have
used a set of refinement levels of size $L^{(n)} = R_{_{\rm AH}}(t=0)
\times 2^{(9-n)}$ so that the finest grid ($n=8$) would initially be
twice the size of the apparent horizon.  This refinement structure
follows the motion of the black holes across the grid and, as the
coordinate growth of the MOTS takes place, we adjust our
grid-structure by first dropping the finest and later the second finest
refinement level.  After dropping the second finest level, we re-adjust
our grid-structure by setting $L^{(n)} = R_{_{\rm AH}}(t) \times
2^{(7-n)}, n = 0,\, \ldots,\, 6$, keeping the ratio $L^{(n)} / R_{_{\rm
    AH}}(t)$ constant as the MOTS grows further.

\begin{figure}
\begin{center}
\includegraphics[width=1.0\textwidth]{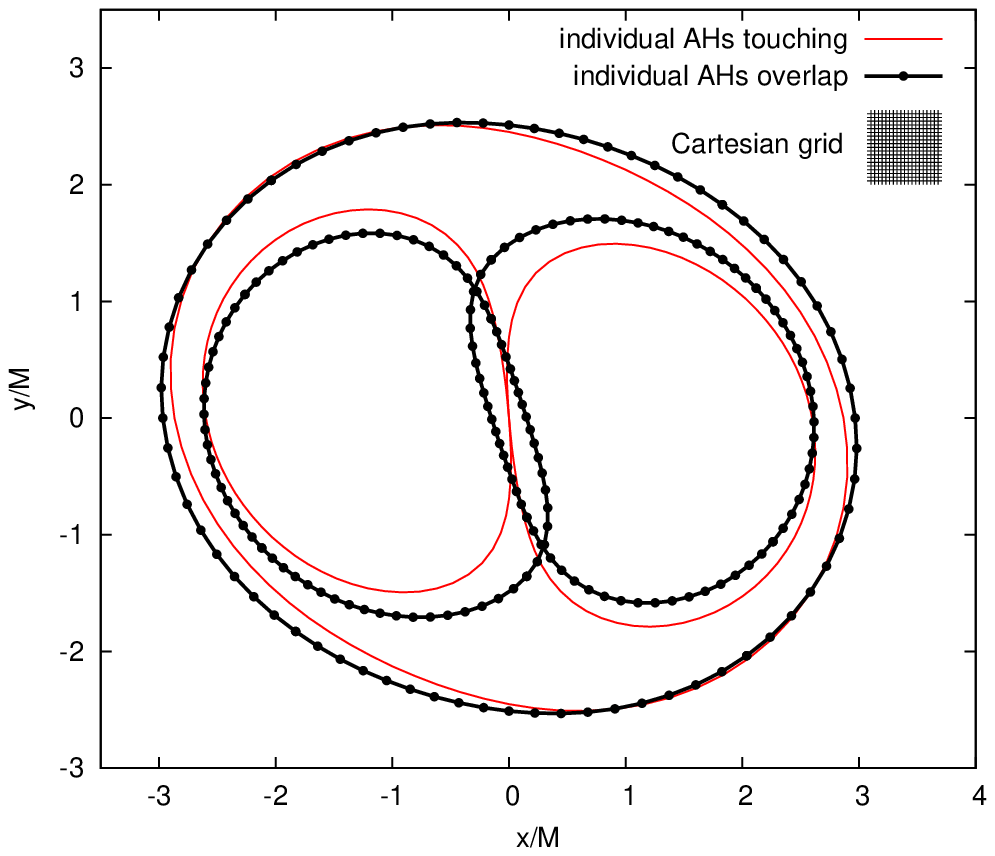}
\end{center}
\caption{Intersection of the MOTS with the $z=0$ plane at
	the time when the individual MOTS first touch
	($t\approx 10.73M$) and just before the individual MOTS
	are lost ($t\approx 11.47M$); for the latter time
	the individual horizon-finder angular grid points
	are also shown.  At the times shown here the finest
	grid has $h^{(6)}=0.032 M$; this is illustrated in the
	plot legend.  Notice that the MOTSs are well-resolved on both
	the Cartesian and angular grids at all times, and that
	they clearly overlap at the latter time.}
\label{fig-qc0-AHsequence}
\end{figure}

At the time of the formation of the common apparent horizon the finest
grid has $h^{(6)}=0.032 M$.  The individual MOTS were tracked as long
as possible.  Eventually the horizon finder algorithm hits the
excision domain of the other black-hole, at which time the individual
MOTS are lost.

Figures~\ref{fig-qc0-PairOfPants-3D} and~\ref{fig-qc0-AHsequence} reveal
an interesting feature of the MOTS dynamics following the formation of a
common apparent horizon. The individual MOTS continue to inspiral until
they touch (apparently at a single point) and then overlap. Note that, by
definition, at the time of touch (and of later intersection) of the
individual MOTS, there already exists a third, common MOTS, which is now
the apparent horizon.  We measure a coordinate velocity of $v \approx
0.16$ of the MOTS as they approach each other. This phenomena, including
the approximate value of $v$, was reproduced using a variety of numerical
parameters, including varying resolutions, regridding rules, and outer
boundary locations. Finer grids allow longer tracking by the apparent
horizon finder, which in turn leads to larger overlaps between the
individual horizons. The figures we present here were obtained with our
best-resolved run.

We have measured the convergence of our QC-0 test by comparing results
obtained at three different resolutions.  At the time of the overlap of
the MOTS, the grid-steps were $h_{\rm h} = 0.032 M, h_{\rm m} = 0.04 M$
and $h_{\rm l} = 0.044 M$.
The MOTS touching time for the three resolutions was
$T_{\rm h} \approx 10.73 M, T_{\rm m} \approx 10.36 M$ and
$T_{\rm l} = 10.29$.
As a first measure of the numerical error, we checked the convergence of
$||C^0||_\infty$ at $t \approx 10.956$. With the excision being a major
  source of numerical error, the $L_\infty$ norm is a reflection of the
  constraint error near the excised points, including those within the
  region of overlapping MOTS. Overall, we measure that
\begin{equation}
\frac{||C^0_{\rm l}||_\infty}{||C^0_{\rm h}||_\infty}
\approx \left( \frac{h_{\rm l}}{h_{\rm h}} \right)^{2.25} .
\end{equation}
Such a convergence order can be explained by bearing in mind that the
derivatives of the metric contained in the constraint $C^0$ are computed
using five-point centered stencils. Near the excision boundary, however,
two of those five points are updated by extrapolation, using the
available three points. This effectively implies that the quantity
$C^0$ is computed near the excision boundary using a three-point
non-centered, first-order derivative stencil, which has an error of
${\cal O}(h^2)$.

As an additional measure of the convergence order we have computed the
$L_2$ norm of the differences between the coordinate-shapes ${\cal S}$ of
the MOTS for the different gridsizes. Given the fact that the three runs
had shape information available at different discrete time values, first
we interpolated in time, from all three sets of data, to a common set
of time-slices
\begin{equation}
t_n = t_{\rm min} + \frac{n-1}{N-1} \times ( t_{\rm max} - t_{\rm min} ),
\quad n = 1 \cdots N .
\end{equation}
The value of $N$ was not critical in this convergence test, given the
smooth behavior of the surfaces. We used  $N=1000$ for the individual MOTS
and $N=100$ for the common apparent horizon.

At each time-level $t_n$, for each resolution, we had a set of grid-points
labeled by their angular grid-index $[I,J]$ as well as Cartesian coordinates.
(Note that for convenience we used the same angular grid-resolution
of the horizon-finder algorithm for all our runs.)
For a given choice of $t_{\rm min},  t_{\rm max}$ and $N$ we define
the $L_2$ norm of the error in the shape of the MOTS to be
\begin{equation}
{ || {\cal S}_l - {\cal S}_h ||_2} = 
\sqrt{
\sum_{I,J,n}
\sum_{i=1}^3 
\left[ (x^i_l(t_n,I,J) - x^i_h(t_n,I,J) \right]^2
\Big / (N_I N_J N)
} \quad . 
\end{equation}

We set $t_{\rm max} = 11.0$ based on
the maximum physical time for which horizon data was available
for the coarsest run. The convergence rate for the individual MOTS 
(with $t_{\rm min} = 0$) was
\begin{equation}
\frac{ || {\cal S}_l - {\cal S}_h ||_2 }{ || {\cal S}_m - {\cal S}_h ||_2 }
\approx \left( \frac{h_{\rm l}}{h_{\rm m}} \right)^{2.42} .
\end{equation}
The accuracy of the touching and overlapping MOTS was measured by
restricting the $L_2$ norm to $t \geq t_{\rm min} = 10.4$.
The measured convergence rate is ${\cal O}(h^{2.59})$, while the
common (apparent) horizon shape was convergent to ${\cal O}(h^{2.57})$.

Overall the accuracy of our QC-0 test is between second and third-order,
which is lower than what was measured for the case of an isolated
Schwarzschild spacetime.  As mentioned above, the most likely causes of
the larger error are the close proximity of the moving excision algorithm
to the apparent horizon finder interpolation stencils and the ${\cal
  O}(dt^3)$ time-interpolation in the mesh refinement algorithm.
(Note that, for time-independent test-cases such as a Schwarzschild
black-hole,
the time-interpolation error will effectively not be seen
due to the small coefficient in front of the corresponding ${\cal
  O}(dt^3)$ error term.)
 
New insight into the peculiar features of apparent horizons have been
revealed in recent numerical simulations~\cite{Schnetter-Krishnan-2005}
and some mathematical machinery has been developed to deal with their
properties in a rigorous way~\cite{Ashtekar05,ams05}. In particular, a recent
theorem due to L. Andersson and J. Metzger (L. Andersson, private
communication) requires that an outer common horizon must already exist
if two MOTS come into contact. It is reassuring and
stimulating that our results are consistent with this theorem.

\section{Conclusions}
\label{sect-discussion}

Over the last few years substantial progress has been made in developing
mathematical theorems which establish the well-posedness of the harmonic
initial-boundary value problem and the stability of its finite difference
approximations. We have incorporated some of this theory in developing an
explicit in time, finite-difference harmonic code and for which we have
presented a series of tests assessing its validity and accuracy. Such tests
range from the evolution of isolated black holes to the head-on collision of
two black holes and over to a binary black hole inspiral and merger. All of
them indicate that stable, convergent and accurate solutions of the Einstein
equations in fully nonlinear regimes can be carried out. Furthermore, the
merger simulations have also revealed that individual MOTS can
touch and even intersect. This novel feature in the dynamics of the MOTS
was not found before but is consistent with theorems on the
properties of MOTS. This finding raises new questions about the
mathematical features of MOTS, which we look forward to exploring.

Overall, these initial results are an important indication that the AEI
harmonic code is capable of contributing to the physical understanding of
binary black holes currently being achieved by numerical simulation. A key
assurance of the validity of numerical results for binary black holes is their
confirmation by a variety of codes based upon different theoretical
formulations and numerical methods.

The work reported here also suggests that there is room for a great deal
of improvement in the numerical methods we have adopted. We have already
emphasized the need for implementing a constraint-preserving treatment of
the outer boundary. In addition, we have carried out very little
exploration to optimize the use of constraint adjustments, the blending
of the subluminal and superluminal algorithm, the extrapolation scheme at
the extraction boundary, mesh refinement, the addition of numerical
dissipation, the choice of gauge conditions, etc. It is particularly
reassuring that the code remains stable even though a large amount of
error is being generated at the jagged excision boundary. It appears that
the outflow nature of the boundary, \textit{i.e.}, its spacelike geometry
combined with the lightlike direction of all characteristics of the
system, leads to an equilibrium between noise generation and its flux out
of the grid.

A number of design choices of our code are based on both the experience
gained while building the Abigel code and on the published work of Pretorius.
It is an optimistic sign that our present results could be
obtained by using what was in a number of cases our initial choice for
the various code details. We regard this as evidence of the fundamental
robustness of the harmonic formulation.

\section*{Acknowledgments}

We thank Lars Andersson for discussing his results concerning unstable
MOTS prior to publication. We also thank Ralf K\"{a}hler
for his help in producing the spacetime rendering of the MOTS
geometry. The numerical results presented in this paper were
obtained on a variety of computer systems, most notably with the
\program{Peyote} and \program{Belladonna} clusters at the AEI. This
work was partially supported by the National Science Foundation under
grant PH-0553597 to the University of Pittsburgh and the SFB-TR7
``Gravitationswellenastronomie'' of the German DFG .

\appendix

\section{Blended subluminal-superluminal evolution}
\label{app-summation-by-parts}

The evolution system (\ref{eqn-harmonic-evolution/2nd-order-in-time}) consists
of coupled quasilinear wave equations whose numerical stability is determined
by the principle part. By the principle of frozen coefficients~\cite{Kreiss89},
the stability analysis can be reduced to a consideration of the wave equation
with shift. Although finite difference approximations to the wave equation is a
well studied problem, the complications introduced by a non-zero shift are
peculiar to the black hole excision problem. This was first recognized
in~\cite{Alcubierre94a}, where it was suggested that the superluminal shift
introduced by tracking the excision boundary could be treated by implicit
methods.

Subsequent studies established the stability of explicit
finite-difference algorithms, with second order accuracy, for the case of
superluminal evolution.  This was first achieved for the 1D wave equation
with shift
\begin{equation}
     g^{tt} \partial_t^2 \Phi + 2 \partial_t \partial_x g^{xt} \Phi +
          g^{xx} \partial_x^2 \Phi = 0,
\label{eq:1dwave}
\end{equation}
in work by
Calabrese~\cite{Calabrese:2005sec} and Szil\'{a}gyi~\etal~\cite{Szilagyi05}.
The standard choice of energy for this system,
\begin{equation}
     E^{(t)}= \frac{1}{2} \int dx \left[
       ( - g^{tt} (\partial_t \Phi)^2 + g^{xx}(\partial_x \Phi )^2
          \right]   ,
\end{equation}
gives rise to a norm when the evolution direction $\partial_t$ is timelike. In
that subluminal case, summation by parts (SBP) can be used to establish
stability of the semi-discrete approximation
\begin{equation}
     g^{tt}\partial_t^2 \Phi  + 2 g^{xt} D_0\partial_t \Phi +
                 g^{xx} D_+ D_- \Phi = 0,
\label{eq:1dsubwave}
\end{equation}
where $D_+$, $D_-$ and $D_0$ are, respectively, the standard forward, backward
and centered finite difference approximations for $\partial_x$. This ensures
that the numerical error is controlled by an estimate for the semi-discrete
version of the energy norm $E^{(t)}$. For most method of lines time
integrators, e.g. Runge-Kutta, this estimate extends to the fully discretized
system. The algorithm (\ref{eq:1dsubwave}) has been extended to the 3D
subluminal case to give a stable SBP boundary treatment~\cite{Babiuc:2006wk}.

However, the algorithm (\ref{eq:1dsubwave}) is unstable (and cannot be
stabilized by Kreiss-Oliger type dissipation) when the evolution is
superluminal, \textit{i.e.}, when the shift is large enough so that
$\partial_t$ is spacelike and
\begin{equation}
    g^{xx}=h^{xx}+ \frac {(g^{xt})^2}{g^{tt}} < 0,
\end{equation}
where $h^{xx}>0$ is the inverse to the spatial metric of the $t=\text{const}$
Cauchy hypersurfaces. In that case, when the energy $E^{(t)}$ is no
longer a norm,
stability can be based upon the positive energy associated with the
time-like normal $n^\mu$ to the Cauchy hypersurfaces,
\begin{equation}
      E^{(n)} = \frac{1}{2} \int dx
     \left[ - \frac{1}{g^{tt}}(g^{tt}\partial_t \Phi +g^{tx} \partial_x \Phi)^2
           + h^{xx} \partial_x^2 \Phi \right].
\end{equation}
As discussed in the 1D case~\cite{Calabrese:2005sec,Szilagyi05}, the
discretization
\begin{equation}
     g^{tt}( \partial_t +  \frac{g^{xt}}{g^{tt}} D_0)^2 \Phi
                + h^{xx} D_+ D_- \Phi = 0.
\label{eq:1dsupwave}
\end{equation}
yields a stable second-order accurate superluminal algorithm.  Stable
superluminal evolution algorithms for the 3D case have been given by
Motamed~\etal~\cite{Motamed06}, where the global stability of a model
black hole excision problem is treated.

Although a stable boundary treatment for the superluminal algorithm
(\ref{eq:1dsupwave}) has been proposed~\cite{Calabrese:2005fp}, its extended
stencil (due to the $D_0^2$ operator) makes this complicated and an SBP
boundary version has not yet been formulated. For this reason we use the 3D
version of the subluminal algorithm (\ref{eq:1dsubwave}) in the neighborhood of
the outer boundary and blend it to the superluminal algorithm
(\ref{eq:1dsupwave}) by introducing the vector
\begin{equation}
  \hat n^{\mu} = (g^{tt}, w g^{it})
\label{eq:nhatdef}
\end{equation}
and the evolution variable
\begin{equation}
  \hat Q = \hat n^{\mu} \partial_\mu \Phi,
\end{equation}
where $w(x^i)$ is a spherically symmetric smooth blending function, with
$w=0$ near the outer boundary and $w=1$ (so that $\hat n^{\mu}=n^\mu$) in
the interior. It suffices to discuss the frozen coefficient case in which
the 1D wave equation (\ref{eq:1dwave}) gives rise to the evolution system
for $\hat Q$ and $\Phi$,
\begin{eqnarray}
    g^{tt} \partial_t \hat Q &=&
       -( 2g^{xt}-\hat n^x) \partial_x(Q - \hat n^x  \partial_x \Phi)
                 -  g^{tt} g^{xx}  \partial_x^2 \Phi
             \nonumber \\
 g^{tt}\partial_t \Phi &=& \hat Q  - \hat n^x \partial_x \Phi  .
\label{eq:Q1}
\end{eqnarray}
Note that introduction of the auxiliary variable $\hat Q$, which reduces
the system to first-order in time, introduces no associated constraints.

For a second-order accurate approximation, we discretize (\ref{eq:Q1})
according to
\begin{eqnarray}
&\hskip -1.0cm
    g^{tt} \partial_t \hat Q =
       -( 2g^{xt}-\hat n^x)D_0  Q  \nonumber 
        + ( 2g^{xt}-\hat n^x)\hat n^x D_+ D_- \Phi
                 -  g^{tt} g^{xx}  D_+ D_- \Phi
             \nonumber \\
&\hskip -1.0cm
 g^{tt}\partial_t \Phi = \hat Q  - \hat n^x D_0 \Phi  .
\label{eq:Q2}
\end{eqnarray}
In the neighborhood of the outer boundary, this reduces to the subluminal
algorithm (\ref{eq:1dsubwave}) and in the interior where $w=1$ it reduces
to the superluminal algorithm (\ref{eq:1dsupwave}). The harmonic code
uses a fourth-order accurate version of (\ref{eq:Q2}) in the interior
region. An alternative scheme for switching between stable subluminal and
superluminal algorithms across the ``artificial horizon'' where $\det
(g^{ij})=0$ is given in~\cite{Motamed06}.

\section*{References}
\bibliography{references}

\begin{thebibliography}{10}

\bibitem{Detweiler-in-Smarr79}
S.~L. Detweiler.
\newblock Black holes and gravitational waves: Perturbation analysis.
\newblock In L.~Smarr, editor, {\em Sources of gravitational radiation}, pages
  211--230. Cambridge University Press, Cambridge, England, 1979.

\bibitem{Hahn64}
Susan~G. Hahn and Richard~W. Lindquist.
\newblock The two body problem in geometrodynamics.
\newblock {\em Ann. Phys.}, 29:304--331, 1964.

\bibitem{Cadez71}
A.~{\v{C}}ade{\v{z}}.
\newblock {\em Colliding Black Holes}.
\newblock PhD thesis, University of North Carolina at Chapel Hill, Chapel Hill,
  North Carolina, 1971.

\bibitem{Eppley75}
Kenneth~R. Eppley.
\newblock {\em The numerical evolution of the collision of two black holes}.
\newblock PhD thesis, Princeton University, Princeton, New Jersey, 1975.

\bibitem{Smarr76}
Larry Smarr, Andrej {\v{C}}ade{\v{z}}, Bryce DeWitt, and Kenneth~R. Eppley.
\newblock Collision of two black holes: Theoretical framework.
\newblock {\em Phys. Rev. D}, 14(10):2443--2452, 1976.

\bibitem{Smarr77}
L.~Smarr.
\newblock Spacetimes generated by computers: Black holes with gravitational
  radiation.
\newblock {\em Ann. N. Y. Acad. Sci.}, 302:569--604, 1977.

\bibitem{Smarr79}
L.~Smarr.
\newblock Gauge conditions, radiation formulae and the two black hole
  collision.
\newblock In L.~Smarr, editor, {\em Sources of gravitational radiation}, page
  245. Cambridge University Press, Cambridge, England, 1979.

\bibitem{Arnowitt62}
Richard Arnowitt, Stanley Deser, and Charles~W. Misner.
\newblock The dynamics of general relativity.
\newblock In L.~Witten, editor, {\em Gravitation: An introduction to current
  research}, pages 227--265. John Wiley, New York, 1962.

\bibitem{Frittelli:2000uj}
Simonetta Frittelli and Roberto G{\'o}mez.
\newblock Ill-posedness in the {E}instein equations.
\newblock {\em J. Math. Phys.}, 41:5535--5549, 2000.

\bibitem{Cook97a}
Gregory~B. Cook \textit{et al.}
\newblock Boosted three-dimensional black-hole evolutions with singularity
  excision.
\newblock {\em Phys. Rev. Lett.}, 80:2512--2516, 1998.

\bibitem{Pretorius:2005gq}
Frans Pretorius.
\newblock Evolution of binary black hole spacetimes.
\newblock {\em Phys. Rev. Lett.}, 95:121101, 2005.

\bibitem{Diener-etal-2006a}
Peter Diener, Frank Herrmann, Denis Pollney, Erik Schnetter, Edward Seidel,
  Ryoji Takahashi, Jonathan Thornburg, and Jason Ventrella.
\newblock Accurate evolution of orbiting binary black holes.
\newblock {\em Phys. Rev. Lett.}, 96(12):121101, March 30 2006.

\bibitem{Campanelli:2005dd}
Manuela Campanelli, Carlos~O. Lousto, Pedro Marronetti, and Yosef Zlochower.
\newblock Accurate evolutions of orbiting black-hole binaries without excision.
\newblock {\em Phys. Rev. Lett.}, 96:111101, 2006.

\bibitem{Baker05a}
John~G. Baker, Joan Centrella, Dae-Il Choi, Michael Koppitz, and James van
  Meter.
\newblock Gravitational wave extraction from an inspiraling configuration of
  merging black holes.
\newblock {\em Phys. Rev. Lett.}, 96:111102, 2006.

\bibitem{Gonzales06tr}
Jo{\'s}e~A. Gonz{\'a}les, Ulrich Sperhake, Bernd Br{\"u}gmann, Mark Hannam, and
  Sascha Husa.
\newblock Total recoil: the maximum kick from nonspinning black-hole binary
  inspiral.
\newblock gr-qc/0610154, 2006.

\bibitem{Herrmann:2006ks}
Frank Herrmann, Deirdre Shoemaker, and Pablo Laguna.
\newblock Unequal-mass binary black hole inspirals.
\newblock gr-qc/0601026, 2006.

\bibitem{deDonder1921}
T.~de~Donder.
\newblock {\em La Gravifique {E}insteinienne}.
\newblock Gauthiers-Villars, Paris, 1921.

\bibitem{Fock-1959}
V.~Fock.
\newblock {\em The Theory of Space, Time, and Gravitation}.
\newblock Pergamon, New York, 1959.

\bibitem{Choquet83}
Y.~Choquet-Bruhat and T.~Ruggeri.
\newblock Hyperbolicity of the 3+1 system of {E}instein equations.
\newblock {\em Comm. Math. Phys}, 89:269--275, 1983.

\bibitem{Garfinkle02}
David Garfinkle.
\newblock Harmonic coordinate method for simulating generic singularities.
\newblock {\em Phys. Rev. D}, 65:044029, 2002.

\bibitem{Szilagyi02b}
B.~Szil{\'a}gyi, B.~Schmidt, and Jeffrey Winicour.
\newblock Boundary conditions in linearized harmonic gravity.
\newblock {\em Phys. Rev. D}, 65:064015, 2002.

\bibitem{Szilagyi02a}
B.~Szil{\'a}gyi and Jeffrey Winicour.
\newblock Well-posed initial-boundary evolution in general relativity.
\newblock {\em Phys. Rev. D}, 68:041501, 2003.

\bibitem{Babiuc:2006wk}
Maria~C. Babiuc, B{\'e}la Szil{\'a}gyi, and J.Winicour.
\newblock Harmonic initial-boundary evolution in general relativity.
\newblock {\em Phys. Rev. D}, 73:064017, 2006.

\bibitem{Babiuc:2006wp}
Maria~C. Babiuc and J.Winicour.
\newblock Constraint-preserving sommerfeld conditions for the harmonic einstein
  equations.
\newblock gr-qc/0612051, 2006.

\bibitem{Babiuc-etal-2005}
Maria~C. Babiuc, B{\'e}la Szil{\'a}gyi, and Jeffrey Winicour.
\newblock Testing numerical relativity with the shifted gauge wave.
\newblock {\em Class. Quantum Grav.}, 23:S319--S342, 2006.

\bibitem{Pretorius:2004jg}
Frans Pretorius.
\newblock Numerical relativity using a generalized harmonic decomposition.
\newblock {\em Class. Quantum Grav.}, 22:425--452, 2005.

\bibitem{Pretorius:2006tp}
Frans Pretorius.
\newblock Simulation of binary black hole spacetimes with a harmonic evolution
  scheme.
\newblock {\em Class. Quantum Grav.}, 23:S529--S552, 2006.

\bibitem{Lindblom:2005qh}
Lee Lindblom, Mark~A. Scheel, Lawrence~E. Kidder, Robert Owen, and Oliver
  Rinne.
\newblock A new generalized harmonic evolution system.
\newblock {\em Class. Quantum Grav.}, 23:S447--S462, 2006.

\bibitem{Scheel-etal-2006:dual-frame}
Mark~A. Scheel, Harald~P. Pfeiffer, Lee Lindblom, Lawrence~E. Kidder, Oliver
  Rinne, and Saul~A. Teukolsky.
\newblock Solving {E}instein's equations with dual coordinate frames.
\newblock gr-qc/0607056, 2006.

\bibitem{Palenzuela-etal-2006-boson-stars}
C.~Palenzuela, I.~Olabarrieta, L.~Lehner, , and S.~Liebling.
\newblock Head-on collisions of boson stars.
\newblock gr-qc/0612067, 2006.

\bibitem{Bona:2002fq}
C.~Bona, Ledvinka. T., and C.~Palenzuela.
\newblock A 3+1 covariant suite of numerical relativity evolution systems.
\newblock {\em Phys. Rev. D}, 66:084013, 2002.

\bibitem{Bona:2004ky}
C.~Bona, T.~Ledvinka, C.~Palenzuela-Luque, and M.~Zacek.
\newblock Constraint-preserving boundary conditions in the {Z4} numerical
  relativity formalism.
\newblock {\em Class. Quantum Grav.}, 22:2615--2634, 2005.

\bibitem{Thornburg95}
Jonathan Thornburg.
\newblock Finding apparent horizons in numerical relativity.
\newblock {\em Phys. Rev. D}, 54(8):4899--4918, October 15 1996.

\bibitem{Thornburg2003:AH-finding}
Jonathan Thornburg.
\newblock A fast apparent-horizon finder for 3-dimensional {C}artesian grids in
  numerical relativity.
\newblock {\em Class. Quantum Grav.}, 21(2):743--766, 21 January 2004.

\bibitem{Schnetter-etal-03b}
Erik Schnetter, Scott~H. Hawley, and Ian Hawke.
\newblock Evolutions in {3D} numerical relativity using fixed mesh refinement.
\newblock {\em Class. Quantum Grav.}, 21(6):1465--1488, 21 March 2004.

\bibitem{Cactusweb}
Cactus~Computational Toolkit.
\newblock \texttt{http://www.cactuscode.org}.

\bibitem{Friedrich96}
Helmut Friedrich.
\newblock Hyperbolic reductions for {E}instein's equations.
\newblock {\em Class. Quantum Grav.}, 13:1451--1469, 1996.

\bibitem{Gundlach2005:constraint-damping}
Carsten Gundlach, Jose~M. Martin-Garcia, G.~Calabrese, and I.~Hinder.
\newblock Constraint damping in the {Z4} formulation and harmonic gauge.
\newblock {\em Class. Quantum Grav.}, 22:3767--3774, 2005.

\bibitem{Calabrese:2005sec}
Gioel Calabrese.
\newblock Finite differencing second order systems describing black hole
  spacetimes.
\newblock {\em Phys. Rev. D}, 71:027501, 2005.

\bibitem{Szilagyi05}
B.~Szil{\'a}gyi, H-O. Kreiss, and J.~Winicour.
\newblock Modeling the black hole excision problem.
\newblock {\em Phys. Rev. D}, 71:104035, 2005.

\bibitem{Calabrese:2005fp}
Gioel Calabrese and Carsten Gundlach.
\newblock Discrete boundary treatment for the shifted wave equation.
\newblock gr-qc/0509119, 2005.

\bibitem{Motamed06}
Mohammad Motamed, M.~C. Babiuc, B.~Szil\'{a}gyi, H-O. Kreiss, and J.Winicour.
\newblock Finite difference schemes for second order systems describing black
  holes.
\newblock {\em Phys. Rev. D}, 73:124008, 2006.

\bibitem{Friedrich99}
Helmut Friedrich and Gabriel Nagy.
\newblock The initial boundary value problem for {E}instein's vacuum field
  equations.
\newblock {\em Commun. Math. Phys.}, 201:619--655, 1999.

\bibitem{Kreiss-Winicour-2002}
Heinz-Otto Kreiss and Jeffrey Winicour.
\newblock Problems which are well-posed in a generalized sense with
  applications to the {E}instein equations.
\newblock {\em Class. Quantum Grav.}, 23:S405--S420, 2006.

\bibitem{Mattsson-Nordstrom-2005:SBP-operators}
Ken Mattsson and Jan Nordstr{\"o}m.
\newblock Summation by parts operators for finite difference approximations of
  second derivatives.
\newblock {\em Journal of Computational Physics}, 199(2):503--540, 20 September
  2004.

\bibitem{Kreiss04a}
Heinz-Otto Kreiss and N.~Anders Petersson.
\newblock A second order accurate embedded boundary method for the wave
  equation with {D}irichlet data.
\newblock {\em SIAM J. Sci. Comput.}, page~31, 2004.

\bibitem{Baker:2002qf}
John Baker, Manuela Campanelli, Carlos~O. Lousto, and Ryoji Takahashi.
\newblock Modeling gravitational radiation from coalescing binary black holes.
\newblock {\em Phys. Rev. D}, 65:124012, 2002.

\bibitem{Ansorg:2005bp}
Marcus Ansorg.
\newblock A double-domain spectral method for black hole excision data.
\newblock {\em Phys. Rev. D}, 72:024018, 2005.

\bibitem{Schnetter-Krishnan-2005}
Erik Schnetter and Badri Krishnan.
\newblock Non-symmetric trapped surfaces in the {S}chwarzschild and {V}aidya
  spacetimes.
\newblock {\em Phys. Rev. D}, 73:021502(R), 2006.

\bibitem{Ashtekar05}
Abhay Ashtekar and Greg Galloway.
\newblock Some uniqueness results for dynamical horizons.
\newblock {\em Advances in Theoretical and Mathematical Physics}, 9(1):1--30,
  2005.

\bibitem{ams05}
L.~Andersson, M.~Mars, and W.~Simon.
\newblock Local existence of dynamical and trapping horizons.
\newblock {\em Phys. Rev. Lett.}, 95:111102, 2005.

\bibitem{Kreiss89}
Heinz~Otto Kreiss and J.~Lorenz.
\newblock {\em Initial-Boundary Value Problems and the Navier-Stokes
  Equations}.
\newblock Academic Press, New York, 1989.

\bibitem{Alcubierre94a}
Miguel Alcubierre and Bernard Schutz.
\newblock Time--symmetric {ADI} and causal reconnection: Stable numerical
  techniques for hyperbolic systems on moving grids.
\newblock {\em J. Comput. Phys.}, 112:44, 1994.

\end{thebibliography}

\end{document}